\documentstyle[aps]{revtex}

\begin{document}

\draft

\title{Transfer/Breakup Modes in the $^{6}$He~+~$^{209}$Bi Reaction Near
and Below the Coulomb Barrier}

\author{
E.F.\,Aguilera$^{1}$, J.J.\,Kolata$^2$, F.M.\,Nunes$^{3,4}$, F.D.\,Becchetti$^5$, P.A.\,DeYoung$^{6}$, M.\,Goupell$^{6}$, V.\,Guimar\~aes$^2$, B.\,Hughey$^6$, M.Y.\,Lee$^5$, D.\,Lizcano$^{1}$, E.\,Martinez-Quiroz$^{1}$, A.\,Nowlin$^6$, T.W.\,O'Donnell$^5$, G.F.\,Peaslee$^7$, D.\,Peterson$^2$, P.\,Santi$^2$, and R.\,White-Stevens$^2$ }

\address{
$^1$ Departamento del Acelerador, Instituto Nacional de Investigaciones Nucleares, A.P. 18-1027, C.P. 11801, Mexico, D.F.\\
$^2$ Physics Department, University of Notre Dame, Notre Dame, Indiana 46556\\
$^3$ Centro Multidisciplinar de Astrof\'isica, Instituto Superior T\'ecnico, 1096 Lisboa-Codex, Portugal\\
$^4$ Department of Science and Technology, Universidade Fernando Pessoa, 4200 Porto, Portugal\\
$^5$ Physics Department, University of Michigan, Ann Arbor, Michigan 48109\\
$^6$ Physics Department, Hope College, Holland, MI 49422\\
$^7$ Chemistry Department, Hope College, Holland, MI 49422\\
}
\date{\today}

\maketitle

\begin{abstract}

Reaction products from the interaction of $^{6}$He with $^{209}$Bi have been measured at energies near the Coulomb barrier.  A $^{4}$He group of remarkable
intensity, which dominates the total reaction cross section, has been observed.  The angular distribution of the group suggests that it results primarily from
a direct nuclear process.  It is likely that this transfer/breakup channel is the doorway state that accounts for the previously observed large sub-barrier
fusion enhancement in this system.

\end{abstract}


A recent investigation of near-barrier and sub-barrier fusion of the
exotic ``Borromean" \cite{zhu93} nucleus $^{6}$He on a $^{209}$Bi target revealed a striking enhancement of the fusion cross section, corresponding to a 25\% reduction in the nominal fusion barrier \cite{kol98}.  Lowering of
the barrier by such an extreme amount is, in fact, a general feature of theoretical predictions for the fusion of the ``neutron halo" nucleus $^{11}$Li
\cite{das96,tak93,hus92,hus95,das94}.  However, the leading effect in
these calculations is purely static and results from the very extended
radius of the valence neutron wave function in $^{11}$Li, which allows
the attractive nuclear force to act at longer distances.  The
two-neutron separation energy for $^{6}$He is considerably larger than
that of $^{11}$Li (0.98 MeV vs. 0.30 MeV), and the valence neutrons are
primarily in a {\it 1p} state rather than a {\it 2s} state and so
experience an angular momentum barrier.  For these reasons, the $^{6}$He
valence neutron wave function does not extend to as large a radius as in $^{11}$Li.  The remarkable suppression of the fusion barrier reported in Ref.\cite{kol98} was therefore somewhat unexpected.  Its origin is investigated in more detail in this work. 

In addition to the static effect, a modest dynamical enhancement of the 
$^{11}$Li fusion cross section was obtained in some of the calculations
\cite{das96,tak93} by coupling to the soft E1 mode.  A similar effect
undoubtedly occurs for $^{6}$He but is unlikely to be the complete
explanation for the observations.  The role played by the projectile
breakup channels, which are possibly important due to the weak 
binding of the valence neutrons, is considerably more controversial.
Several groups \cite{tak93,hus92,hus95} have reported that coupling
to these channels reduces the fusion cross section near the barrier,
while Dasso and Vitturi \cite{das94} predict only enhancement even
in the presence of very strong projectile breakup.  Finally, as
pointed out in Ref.\cite{kol98}, none of the existing calculations include the nucleon-transfer degree of freedom.  It was suggested there that the observed
enhancement may result from coupling to positive Q-value neutron transfer
channels, leading to ``neutron flow" between the projectile and target
as discussed by Stelson, {\it et al.}\cite{stel90}.  In this work, we
report the results of an experiment to measure transfer and/or breakup 
products from the $^{6}$He~+~$^{209}$Bi reaction near and below
the barrier to shed light on the mechanism causing the strong suppression of the fusion barrier in this system.

The $^{6}$He beam used in the experiment was produced by the {\it TwinSol} radioactive nuclear beam (RNB) facility at the University of Notre Dame 
\cite{lee99}.  Two large superconducting solenoids act as thick lenses to 
collect and focus the secondary beam of interest onto a spot that was
typically 5 mm full width at half maximum (FWHM).  In this work, the 
primary beam was $^{7}$Li at an energy of 30.5 MeV, incident on a 
gas target with a 2$\mu$ Havar entrance window.  The cell was 2.5 cm
long and contained He gas at a pressure of 1 atm.  The purpose of the
gas was to cool the exit window, a 12$\mu$ foil of $^{9}$Be in which 
$^{6}$He is produced via the $^{9}$Be($^{7}$Li,$^{6}$He) proton transfer
reaction.  Primary beam currents of up to 300 particle$^.$nA (pnA) were available, resulting in a maximum $^{6}$He rate of 10$^5$ s$^{-1}$.  The
secondary beam flux was calibrated by inserting a Si $\Delta$E-E telescope at the secondary target position and reducing the intensity of the primary beam by three orders of magnitude, so that the $^{6}$He particles could be directly counted while at the same time the primary beam current was measured in a Faraday cup.  The secondary beam was contaminated by ions having the
same magnetic rigidity as the desired $^{6}$He beam.  This contamination
was reduced by placing an 8$\mu$ Havar foil at the crossover point
between the two solenoids.  Differential energy loss then helps to
eliminate unwanted ions from the beam prior to the secondary target,
which was a 3.2 mg/cm$^2$ Bi layer evaporated onto a 100 $\mu g/cm^2$ 
polyethylene backing.  The remaining contaminant ions were identified on an event-by-event basis using time-of-flight (TOF) techniques.  The TOF of the particles was obtained from the time difference between the occurrence of the secondary reaction and the RF timing pulse from a beam buncher.  The time resolution of better than 3 ns (FWHM) was adequate to separate $^6$He from all
contaminants, except for $^3$H which has the same mass-to-charge ratio
and therefore the same velocity as $^6$He.  As discussed below, this
required us to carry out a separate experiment with a $^3$H beam
of the appropriate energy, also produced using {\it TwinSol}.  The
laboratory energy of the $^6$He beam was 22.5 MeV for the above-
barrier measurement.  This was reduced to 19 MeV for the below-barrier
measurement via energy loss in a polyethylene foil.  In both cases, the
energy resolution of the beam was determined to be 1.2 MeV FWHM.  

The reaction events, and also elastically-scattered particles,
were detected with five Si $\Delta$E-E telescopes placed at various angles on either side of the beam.  Each telescope had a circular collimator that subtended a solid angle between 26-48~msr, corresponding to an overall effective angular resolution of 9$^{\rm o}$-11$^{\rm o}$ (FWHM), computed by folding in the acceptance of the collimator with the spot size and angular divergence of the beam.  A typical spectrum, taken at 22.5 MeV and an angle of 150$^{\rm o}$, is shown in Fig. 1.  The elastic $^6$He group is visible, along with $^4$He and H isotopes.  A strong, isolated group of $^4$He ions having a mean energy about
2 MeV less than that of the $^6$He elastic group is clearly visible.
Note that this spectrum is gated by TOF, so that scattered $^4$He ions 
in the secondary beam (which have an energy 1.5 times that of $^6$He) 
have been identified and removed.  The $^4$He ions at lower energy, below 
the isolated peak, come from reactions in the backing of the target, as determined from a separate spectrum taken with a backing foil without Bi.
Also visible in Fig. 1 is a $^3$H group, which as noted above cannot
be identified on the basis of TOF.  This could be a problem since
the $^{209}$Bi($^{3}$H,$^{4}$He) reaction has a large positive Q-value
and the $^4$He ions in the isolated group might be coming from this
reaction.  This possibility was eliminated in a separate
experiment with a $^3$H beam of the appropriate energy (half that of
$^6$He) which showed no events in this region.

Angular distributions obtained for the isolated $^4$He group are illustrated
in Fig. 2.  They are broad and approximately Gaussian in form, with a
centroid that moves backward at the lower energy (Table I).  Their most striking
feature, however, is the very large magnitude of the total cross section,
equal to 773 mb at 22.5 MeV and 643 mb at 19 MeV.  For comparison purposes,
the fusion cross sections measured at these energies (Ref.\cite{kol98})
are 310 mb and 75 mb, respectively.  This initially very surprising
result was confirmed by the elastic-scattering angular distributions
(Fig. 3) which imply total reaction cross sections of
about 1170 mb and 670 mb at the two energies, consistent with the
sum of the fusion and $^4$He yields within experimental error.  
The curves shown in Fig. 3 are obtained from optical-model fits 
to the data, resulting in the
parameters given in Table II.  The imaginary well depth was found to
be about 75\% greater at the lower energy.  This rapid energy dependence
implies that the effective imaginary potential is not well described
by a Woods-Saxon form.  Also shown in Fig. 3 are optical-model predictions
using parameters that are appropriate for $^4$He scattering, but with radius parameters increased to correspond to the larger size of $^6$He.  This illustrates the expectations for elastic scattering of a ``normal" nuclear system near to and below the barrier.  The predicted total reaction
cross sections are 238 mb and 5.2 mb, respectively.

The $^4$He group seen in this experiment dominates the
total reaction cross section at near- and sub-barrier energies.  
It is therefore important to determine the reaction mechanism that 
accounts for this very large yield.  Unfortunately, it is not possible to separate neutron transfer from breakup modes using only the present data.
Based on the absence of events in the appropriate energy region,
two-nucleon transfer to the ground state of $^{209}$Bi (Q = +8.8 MeV) is unimportant.
This was confirmed with a finite-range DWBA calculation for dineutron transfer; the predicted maximum yield was less than 0.1 mb/sr. Here and in the other calculations reported in this work: i) the two neutrons were coupled to
a relative $\it s$ state and their motion relative to both $^4$He and $^{209}$Bi has $\ell$ = 0, ii) the incoming and outgoing distorted waves were defined by the corresponding optical-model parameters given in Table II, and iii) the code FRESCO \cite{fresco} was used and the transfer operator included the remnant term.  

Single neutron transfer
followed by breakup of the remaining unstable $^5$He has a very different 
Q-value, as does direct breakup into $^4$He plus two neutrons.  However,
outgoing $^4$He ions resulting from either of these mechanisms could
be accelerated by the Coulomb field of the target to approximately the energy
of the observed group, so neither process can be eliminated on the
basis of the energy of the outgoing $\alpha$-particle group.  We will
return to this point.

The experimental angular distributions do reveal some information about
the reaction mechanism.  The sideward peaking, and the
fact that the maximum of the distribution shifts to a larger angle
at the lower energy, argue for a nuclear process.  Coulomb dissociation, 
while undoubtedly present, has a forward-peaked angular distribution that cannot 
completely account for the observations.  Nuclear-induced direct breakup,
on the other hand, should play an important role in the
measured angular range.  The result of a coupled-channels calculation of
direct nuclear breakup at 22.5 MeV is illustrated by the thin solid line 
in Fig. 2.  In this calculation, the NN-$\alpha$ interaction was adjusted
to reproduce the features of the known $^{6}$He resonance \cite{res}; the
continuum was included up to 4 MeV above threshold.   The predicted maximum yield is too small by a factor of four, and the angular distribution is somewhat broader than observed.  Part of the discrepancy at forward angles between theory and experiment in Fig. 2 might be due to the contribution from Coulomb breakup.  Breakup calculations including the Coulomb term are much more
difficult to perform because of the very long range of the couplings and full
convergence has not yet been attained.

Another possibility is transfer to excited states in $^{211}$Bi.  We first
assumed $\ell$ = 0 transfer of a dineutron to a ``barely-bound" state with
a binding energy of 0.1 MeV.  The calculated angular distribution is shown as
the dashed line in Fig. 2.  The absolute yield is much too small; the theoretical prediction in Fig. 2 has been multiplied by a factor of ten.  The result of a preliminary nucleon-transfer calculation including continuum states is more encouraging.  In this calculation, the
valence neutron pair in $^6$He was transferred into a range of unbound states
in $^{211}$Bi, up to 8 MeV above threshold.  All couplings between these
states and the $^{209}$Bi ground state were included, and the interaction
in the $^{211}$Bi continuum was assumed to be the same as that which binds
the dineutron in the ground state.  This is the best that can
be done based on the present lack of knowledge of the structure of $^{211}$Bi 
at high excitation.  Under these conditions, the wave function of the valence dineutron is very extended, as there are no Coulomb or angular momentum barriers to be overcome.  Furthermore, since the 
favored ``Q-window" for neutron transfer is at Q $\simeq 0$, $\it{i.e}$, close to the observed maximum in the experimental yield, the reaction gains a kinematic enhancement.  As a result, the predicted cross section is very large, comparable to the experimental yield, and the angular distribution is
characteristic of a nuclear process and appears very similar 
to the dashed curve in Fig. 2.  In addition, coupling to the fusion
channel is included consistently, and the calculation predicts an enhancement in sub-barrier fusion which is comparable to our previous measurement \cite{kol98}.

It is also possible that single neutron transfer, followed
by breakup of the remaining $^{5}$He, could occur.  The
$^{4}$He residue would then be Coulomb accelerated as discussed above.
The states near the Fermi surface all have high angular momentum,
though, so the transfer might be suppressed by an angular momentum
barrier.  In any event, this calculation has not yet been attempted.  
Clearly, much more theoretical work remains to be done before the origin 
of the observed very strong $^{4}$He yield is understood in any detail.

As to the speculation in Ref. \cite{kol98} regarding ``neutron
flow", the observed Q-value spectrum conclusively shows that ground-state transfer, with its corresponding high positive Q value, is unimportant.  However, as discussed above, the positive Q value does play a role in 
making the continuum states in $^{211}$Bi accessible within the preferred Q window.  The transfer to these unbound states could be described as neutron flow, though transfer/breakup seems more appropriate under the circumstances.
Nevertheless, the preliminary coupled-channels calculation does show
that coupling to the transfer/breakup channels has the potential to explain
the large sub-barrier fusion enhancement seen in the $^{6}$He~+~$^{209}$Bi
system.  Apparently it is the strength of the transfer channel and not the positive Q value per se that determines the enhancement, in agreement with the conclusions of Henning, $\it {et~al.}$ \cite{hen87} for ``normal" nuclei.

In conclusion, we have for the first time measured near-barrier and
sub-barrier transfer/breakup yields for an exotic ``Borromean"
nucleus, $^{6}$He, on a $^{209}$Bi target.  An isolated $^{4}$He group was observed at an effective Q-value of approximately -2 MeV.  The integrated cross section for this group is exceptionally large, greatly exceeding the fusion yield both above and below the barrier.  Moreover, simultaneously-measured
elastic scattering angular distributions require total reaction
cross sections that confirm this large yield.  Preliminary coupled-channels
calculations suggest that the corresponding reaction mechanism can best
be described as direct breakup plus neutron transfer to unbound states in $^{211}$Bi.  The latter process is enhanced by the large radial extent of the wave function of the unbound states, leading to excellent overlap with the weakly-bound valence neutron orbitals of $^{6}$He.  It also experiences a kinematic enhancement due to the fact that the large positive ground-state Q value for transfer makes the neutron unbound states accessible within 
the optimum ``Q-window" at Q $\simeq 0$.  The resulting mechanism bears some
resemblance to ``neutron flow" \cite{stel90}, and to the ``neutron
avalanche" discussed by Fukunishi, {\it et al.} \cite{fu93} in the
context of ``neutron skin" nuclei. Finally, the calculations also predict
an enhancement in the sub-barrier fusion yield due to coupling to
the transfer/breakup channel, which strongly suggests that this is the
``doorway state" that accounts for the remarkable suppression of the
fusion barrier observed in a previous experiment\cite{kol98}.

This work was supported by the National Science Foundation under 
Grants No. PHY99-01133, PHY98-70262, PHY98-04869, and PHY97-22604,
and by the CONACYT (Mexico).  One of us  (V.G.)  was financially
supported by FAPESP (Funda\c c\~ao de Amparo a Pesquisa do Estado 
de S\~ao Paulo - Brazil) while on leave from the UNIP (Universidade Paulista).

\vspace{2cm}


\begin{table}[hb]
\label{tab}
\caption[ ]{ Parameters of the Gaussian fits to the data shown in Fig. 2.
}
\begin{tabular}{c c c c}
$E_{lab}~(MeV)$ & $Centroid~(deg)$ & $FWHM~(deg)$ & $\sigma_{total}~(mb)$
\\[0.8mm]
\hline
\\[-3.4mm]
 ~~$22.5$ & ~86.2~(2.5) & 119.6~(~5.6) &  773~(31)\\
 ~~$19.0$ & 116.6~(5.3) & 131.8~(19.7) &  643~(42)\\
\end{tabular}
\end{table}

\begin{table}[hb]
\label{tab}
\caption[ ]{ Optical-model parameters used in the calculations shown in
Fig. 3.  The third row gives a potential determined for $^{4}$He + $^{209}$Bi
at an incident energy of 22.0 MeV \cite{bar74}.  In each case, the Coulomb
radius was taken to be 7.12 fm.
}
\begin{tabular}{c c c c c c c r}
$E_{lab}~(MeV)$ & $V~(MeV)$ & $R~(fm)$ & $a~(fm)$ &
         $W~(MeV)$ & $R_I~(fm)$ & $a_I~(fm)$& $\sigma_{reac}~(mb)$
\\[0.8mm]
\hline
\\[-3.4mm]
 ~~$22.5$ & 150.0 & 7.96 &  0.68 & $27.8~^{a)}$ & 9.38 & 0.99 &  1167~~~~\\
 ~~$19.0$ & 150.0 & 7.96 &  0.68 & $47.8~^{a)}$ & 9.38 & 0.99 &  ~668~~~~\\
 ~~$22.5$ & 100.4 & 8.57 &  0.54 & $44.3~^{b)}$ & 7.12 & 0.40 &  ~238~~~~\\
\end{tabular}
{\footnotesize
$^{a)}$ Volume imaginary potential.
$^{b)}$ Surface imaginary potential.
}
\end{table}
%
%

\vspace{2cm}

\begin{figure}
\caption{A $\Delta$E vs. E$_{\rm TOTAL}$ spectrum taken at 
$\Theta_{\rm LAB}~=~150^{\rm o}$, at a laboratory $^6$He energy of 22.5 MeV.
The energy calibration is 80 keV/channel.}
\end{figure}

\begin{figure}
\caption{Experimental angular distributions for the $^4$He group measured 
  in this work.  The solid curves are Gaussian fits to the data, with the 
parameters given in Table I.  The thin solid curve is the result
of a direct nuclear breakup calculation.  The dashed curve is a calculation
of transfer to a barely bound state; the magnitude of the predicted
yield has been multiplied by a factor of 10 in this case.}
\end{figure}

\begin{figure}
\caption{The experimental elastic-scattering angular distributions.  The
ratio to the Rutherford cross section is compared with optical-model
fits (solid curves), which yield the parameters given in Table II.  The
dashed curves are calculations made with potentials appropriate for $^4$He
+$^{209}$Bi \protect\cite{bar74}, but with a radius appropriate for $^6$He.  
The total reaction cross section computed with this potential at 19.0 MeV
is 5.2 mb.  Reaction cross sections corresponding to the other curves are
given in Table II.
}

\end{figure}

\end{document}